\documentclass[12pt]{article}
\usepackage{graphicx}
\fontsize{14}{24pt}\selectfont

\title{Corrections to the $ns$-levels of hydrogen atom in deformed space with minimal length}
\author{M. M. Stetsko\footnote{E-mail: mykola@ktf.franko.lviv.ua}\
\\
  {\small Department of Theoretical Physics, Ivan Franko National University of Lviv,}\\
{\small 12 Drahomanov Str., Lviv, UA-79005, Ukraine
         }}

\begin{document}
\maketitle

\abstract{We investigated the hydrogen atom problem with deformed
Heisenberg algebra leading to the existence of minimal length.
Using modified perturbation theory developed in our previous work
[M. M. Stetsko and V. M. Tkachuk, Phys. Rev. A {\bf 74}, 012101
(2006)] we calculated the corrections to the arbitrary $s$-levels
for hydrogen atom. We received a simple relation for the
estimation of minimal length. We also compared the estimation of
minimal length obtained here with the results obtained in the
preceding investigations.}

\section{Introduction}
In recent years there has been a growing interest in quantum
mechanical systems with deformed (generalized) commutation
relations. Deformed commutation relations arose as a natural
generalization of the canonical ones. So it is interesting to
consider the quantum mechanical problems where the position and
the momentum operators obey the generalized commutation relations.
Deformed commutation relations appeared in the quantum gravity and
string theory, where it was expected that generalized commutation
relations might eliminate some disadvantages of these theories.
String theory and quantum gravity implied on the existence of
minimal observable length \cite{gross, maggiore, witten}. Such a
suggestion leads to the appropriate deformation of commutation
relations between the momentum and position operators
\cite{kempf1, kempf2, kempf3, kempf4, kempf5}. In the
$D$-dimensional case deformed Heisenberg algebra introducing
minimal length takes the tensorial form:
\begin{eqnarray}\label{algebra}
\begin{array}{l}
[X_i, P_j]=i\hbar(\delta_{ij}(1+\beta P^2)+\beta'P_iP_j),\, [P_i,
P_j]=0,
\\
{[X_i, X_j]}=i\hbar\frac{(2\beta-\beta')+(2\beta+\beta')\beta
P^2}{1+\beta P^2}(P_iX_j-P_jX_i).
\end{array}
\end{eqnarray}
The hydrogen atom is one of the simplest quantum mechanical
systems allowing not only a highly accurate theoretical prediction
but also having a well investigated experimental spectrum
\cite{eides, karshenboim}. So it is interestingly to examine the
hydrogen atom problem when the position and momentum operators
obey deformed commutation relations (\ref{algebra}). Such a
problem was considered for the first time by Brau \cite{brau} in a
particular case when $\beta'=2\beta$. In \cite{benczik} hydrogen
atom was investigated in a general case when $\beta'\neq 2\beta$.
The authors developed the perturbation theory for calculating
corrections to the energy spectrum. But that perturbation theory
gave a possibility to calculate corrections to the energy levels
only if $l\neq 0$. Then for calculating corrections to the
$s$-levels the authors used a numerical method and the cutoff
procedure.

In our previous work \cite{mykola} we developed a modified
perturbation theory that gives a possibility to calculate
corrections for arbitrary energy levels. This problem was
considered in the general case when $\beta'\neq 2\beta$. We also
received an analytical expression for the corrections to the $1s$
and $2s$ energy levels. As was shown these results can be reduced
to the results obtained in \cite{brau} when $\beta'=2\beta$ and
are in good agreement with the results obtained in \cite{benczik}
when $\beta'\neq 2\beta$ and $l\neq 0$.

In the present work we continue to investigate the hydrogen atom
problem in deformed space that leads to the existence of minimal
length. We will consider a general case when $\beta'\neq 2\beta$.
We will calculate the corrections to the arbitrary $s$-levels.

This paper is organized as follows. In the second section we
calculate corrections to the arbitrary $s$-levels using the
perturbation theory proposed in our previous work \cite{mykola}.
In the third section we get a simple relation that gives a
possibility to estimate minimal length. We also compare the
estimation obtained here with the estimations received in the
preceding works \cite{benczik, mykola}. And finally the fourth
section contains the discussion.

\section{Corrections to the energy of $ns$-levels for hydrogen atom}
In this section we consider the eigenvalue problem for hydrogen
atom in a three-dimensional case
\begin{equation}\label{coulomb}
\left(\frac{\textbf{P}^2}{2m}-\frac{e^2}{R}\right)\Psi=E\Psi,
\end{equation}
where the operators of position $X_i$ and momentum $P_i$ obey the
deformed commutation relation (\ref{algebra}) and
$R=\sqrt{\sum^3_{i=1}X^2_i}$.

As was shown in our preceding investigation \cite{mykola} the
following representation satisfies the algebra in the first order
over the parameters $\beta$, $\beta'$.
\begin{eqnarray}\label{rep1}
\left\{
\begin{array}{l}
 X_i=x_i+\frac{2\beta-\beta'}{4}\left(x_ip^2+p^2x_i\right),
\\
P_i=p_i+\frac{\beta'}{2}p_ip^2;
\end{array}
\right.
\end{eqnarray}
where $p^2=\Sigma_{k=1}^3p_k^2$ and operators $x_i$, $p_i$ obey
canonical commutation relations $[x_i, p_j]=i\hbar\delta_{ij}$.
For undeformed  Heisenberg algebra the position representation may
be taken: $x_i=x_i$, $p_i=i\hbar\frac{\partial}{\partial x_i}$.

 In paper \cite{mykola} it was shown that Hamiltonian of hydrogen atom (\ref{coulomb})
 can be written in the linear approximation
over the deformation parameters
\begin{equation}\label{H}
H=\frac{p^2}{2m}+\frac{\beta'p^4}{2m}-e^2\left[\frac{1}{\sqrt{r^2+b^2}}-\frac{2\beta-\beta'}{4}\right.
\left.\left(\frac{1}{r}p^2+p^2\frac{1}{r}\right)\right].
\end{equation}
where $r=\sqrt{\sum^3_{i=1}x^2_i}$.

 As was noted in \cite{mykola} one can calculate
corrections to the arbitrary energy levels of hydrogen atom having
Hamiltonian (\ref{H}). It is necessary to say that for calculation
 corrections to the energy levels with nonzero angular momentum
one can use a somewhat different approach \cite{benczik,mykola}.

We rewrite Hamiltonian (\ref{H}) in the form:
\begin{equation}
H=H_0+V
\end{equation}
where $H_0$ is Hamiltonian of ordinary hydrogen atom and $V$ is
the perturbation caused by deformation.
\begin{equation}
V=\frac{\beta'p^4}{2m}-e^2\left[\frac{1}{\sqrt{r^2+b^2}}-\frac{1}{r}\right.
\left.-\frac{2\beta-\beta'}{4}\left(\frac{1}{r}p^2+p^2\frac{1}{r}\right)\right].
\end{equation}
where $b=\hbar\sqrt{\alpha}$, and $\alpha=2\beta-\beta'$.

So having the eigenfunctions for excited $s$-levels one can
calculate corrections to the energy spectrum. At first we consider
the correction caused by the term $\frac{1}{\sqrt{r^2+b^2}}$. We
have
\begin{equation}\label{NScorr}
\left\langle\Psi_{ns}\left|\frac{1}{\sqrt{r^2+b^2}}\right|\Psi_{ns}\right\rangle=
\frac{\pi}{2}\frac{e^2}{a}\frac{(n-1)!}{[n!]^3}\sum^{n-1}_{i,j=0}\zeta^{2n-i-j}C^i_{n-1}C^i_nC^j_{n-1}C^j_ni!j!
\frac{\partial^{2n-i-j}}{\partial\zeta^{2n-i-j}}\left[H_0(\zeta)-Y_0(\zeta)\right]
\end{equation}
where $\zeta=\frac{2b}{na}$, $a$ is the Bohr radius and $H$, $Y$
are the Struve and the Bessel functions, respectively
\cite{abramowitz}.

Hamiltonian (\ref{H}) contains the terms linear over the
deformation parameters. So we write corrections up to the first
order over $\alpha$ (or $b^2$). As one can see for such an
approximation the leading contribution is given by the derivatives
from the Bessel functions and, conversely, we can neglect the
derivatives from the Struve functions.

So we write the expression ignoring the derivatives from the
Struve functions
\begin{equation}\label{e1}
\left\langle\Psi_{ns}\left|\frac{1}{\sqrt{r^2+b^2}}\right|\Psi_{ns}\right\rangle=
-\frac{\pi}{2}\frac{e^2}{a}\frac{(n-1)!}{[n!]^3}\sum^{n-1}_{i,j=0}\zeta^{2n-i-j}C^i_{n-1}C^i_nC^j_{n-1}C^j_ni!j!
\frac{\partial^{2n-i-j}}{\partial\zeta^{2n-i-j}}Y_0(\zeta).
\end{equation}

The expression (\ref{e1}) can be rewritten in the following way:
\begin{eqnarray}\label{e2}
\left\langle\Psi_{ns}\left|\frac{1}{\sqrt{r^2+b^2}}\right|\Psi_{ns}\right\rangle=
-\frac{\pi}{2}\frac{e^2}{a}\frac{(n-1)!}{[n!]^3}\Big(n^2[(n-1)!]^2\zeta^2
\frac{\partial^2}{\partial\zeta^2}Y_0(\zeta)+ \nonumber
\\
\\
+\mathop{\sum^{n-1}_{i,j=0}}\limits_ {i=j\neq
n-1}\zeta^{2n-i-j}C^i_{n-1}C^i_nC^j_{n-1}C^j_ni!j!
\frac{\partial^{2n-i-j}}{\partial\zeta^{2n-i-j}}Y_0(\zeta)\Big)\nonumber
\end{eqnarray}
The derivatives from the Bessel functions can be represented in
the form:
\begin{equation}\label{deriv}
\frac{\partial^k}{\partial\zeta^k}Y_0(\zeta)=\frac{1}{2^k}\sum^k_{l=0}(-1)^lC^l_kY_{2l-k}(\zeta)
\end{equation}
We use this representation for derivatives of the Bessel functions
and substitute relation (\ref{deriv}) in expression (\ref{e2})
\begin{eqnarray}\label{e3}
\left\langle\Psi_{ns}\left|\frac{1}{\sqrt{r^2+b^2}}\right|\Psi_{ns}\right\rangle=
-\frac{\pi}{2}\frac{e^2}{a}\frac{(n-1)!}{[n!]^3}\left(\frac{n^2[(n-1)!]^2}{2}\zeta^2[Y_2(\zeta)-Y_0(\zeta)]+\right.\nonumber
\\
\\
\left.\mathop{\sum^{n-1}_{i,j=0}}\limits_ {i=j\neq
n-1}C^i_{n-1}C^i_nC^j_{n-1}C^j_ni!j!\left(\frac{\zeta}{2}\right)^{2n-i-j}\sum^{2n-i-j}_{l=0}(-1)^lC^l_{2n-i-j}Y_{2l-2n+i+j}
(\zeta)\right)\nonumber
\end{eqnarray}
 We rewrite the expression (\ref{e3}) so that it should take into account
only two leading terms in the last sum,
\begin{eqnarray}
\begin{array}{c}
\left\langle\Psi_{ns}\left|\frac{1}{\sqrt{r^2+b^2}}\right|\Psi_{ns}\right\rangle=
-\frac{\pi}{2}\frac{e^2}{a}\frac{(n-1)!}{[n!]^3}\left(\frac{n^2[(n-1)!]^2}{2}\zeta^2[Y_2(\zeta)-Y_0(\zeta)]+\right.
\\
\\
\mathop{\sum^{n-1}_{i,j=0}}\limits_ {i=j\neq
n-1}C^i_{n-1}C^i_nC^j_{n-1}C^j_ni!j!\left(\frac{\zeta}{2}\right)^{2n-i-j}2(-1)^{2n-i-j}
\Big[Y_{2n-i-j}(\zeta)-(2n-i-j)Y_{2n-i-j-2}(\zeta)\Big]\Big).
\end{array}
\end{eqnarray}
Then we develop the Bessel functions in the series and take into
consideration only the terms that gives the contributions linear
in deformation parameters. So we have
\begin{eqnarray}\label{e4}
\left\langle\Psi_{ns}\left|\frac{1}{\sqrt{r^2+b^2}}\right|\Psi_{ns}\right\rangle=
\frac{e^2}{a}\frac{(n-1)!}{[n!]^3}\left[\frac{n^2[(n-1)!]^2}{2}\left(\zeta^2
\left[\ln(\frac{\zeta}{2})+\gamma+\frac{1}{2}\right]+2\right)\right.+\nonumber
\\\nonumber
\\
\mathop{\sum^{n-1}_{i,j=0}}\limits_ {i=j\neq
n-1}C^i_{n-1}C^i_nC^j_{n-1}C^j_ni!j!(-1)^{2n-i-j}\left((2n-i-j-1)!+\right.
\\\nonumber
\\
\left.\left.\left[(2n-i-j-2)!+(2n-i-j)(2n-i-j-3)!\frac{\zeta^2}{4}\right]\right)\right]\nonumber
\end{eqnarray}

At last one can calculate the contributions into the energy
spectrum caused by the terms $\frac{1}{r}p^2+p^2\frac{1}{r}$,
$p^4$ and $\frac{1}{r}$. Since these calculations are very simple
we can write corrections to the hydrogen atom spectrum in linear
approximation over the deformation parameters.
\begin{eqnarray}\label{maincorrecion}
\Delta
E^{(1)}_{ns}=\langle\Psi_{ns}|V|\Psi_{ns}\rangle=\frac{e^2\hbar^2}{a^3n^3}\left(\frac{2\beta+\beta'}{l+\frac{1}{2}}
-\frac{\beta+\beta'}{n}\right)+\frac{e^2}{n^2a}-\frac{e^2}{na}-\frac{e^2\hbar^2}{n^3a^3}(2\beta-\beta')\times\nonumber
\\\nonumber
\\
\left(\ln\left(\frac{\hbar^2(2\beta-\beta')}{n^2a^2}\right)+2\gamma+1\right)-\frac{e^2}{a}\frac{(n-1)!}{[n!]^3}
\left(\mathop{\sum^{n-1}_{i,j=0}}\limits_ {i= j\neq
n-1}C^i_{n-1}C^i_nC^j_{n-1}C^j_ni!j!(-1)^{2n-i-j}\times\right.
\\\nonumber
\\
\left.\left[(2n-i-j-1)!+\left ((2n-i-j-2)!+(2n-i-j)(2n-i-j-3)!
\frac{\hbar^2(2\beta-\beta')}{n^2a^2}\right)\right]\right).\nonumber
\end{eqnarray}

It is easy to verify that in the special case $n=1$ and $n=2$ one
can obtain the same corrections as calculated in our previous work
for the $1s$ and $2s$ levels, respectively.

\section{Estimation of minimal length}
Finally we can proceed to the estimation of minimal length. In
\cite{benczik,mykola} it was supposed that minimal length effects
were hidden in the discrepancy between theoretical and
experimental values of Lamb shift for the $s$-levels of hydrogen
atom. But calculations of Lamb shift corrections for hydrogen atom
contain some inaccuracies. These inaccuracies in the determination
of the Lamb shift for arbitrary $s$-levels are caused by the
contributions that take into account proton charge distribution
and yet uncalculated state-independent corrections \cite{eides}.
As was noted in \cite{eides,Jentschura} for the determination of
the Lamb for arbitrary $s$-states it is useful to utilize a
special difference
\begin{equation}\label{difference}
\Delta_n=n^3\Delta E_L(ns)-\Delta E_L(1s).
\end{equation}
 Using such a difference one can avoid
the problems noticed above. Having this difference the minimal
length can be estimated in the similar way as it was shown in
previous investigations \cite{benczik,mykola}. So, for the
estimation of the minimal length we suppose that a shift of energy
levels caused by the deformation of the commutation relations does
not exceed the difference between theoretical and experimental
values for specially introduced expression (\ref{difference}).

 As was shown in the recent work \cite{Jentschura} the calculated value of the difference
(\ref{difference}) for the $1s$ and $2s$ states is in the units of
frequency $\Delta^{\textrm{theor}}_2=187\,225.70\,(5)$ kHz. The
experimentally measured values of the Lamb shift are
$L(1s_{1/2})=8\,172\,840\,(22)$ kHz and
$L(2s_{1/2})=1\,045\,009.4(65)$ kHz for the $1s$ and $2s$ levels,
respectively \cite{Beauvoir}. So we have
$\Delta^{\textrm{exprt}}_2=187\, 235. 2$ kHz. As one can see the
inaccuracy of theoretical prediction is much smaller than the
experimental one.

For a more accurate estimation of the minimal length it is
necessary to use the experimental data of the Lamb shift with
uncertainty of the same order as the theoretical one. As is known
to obtain the experimental values of the Lamb shift for the $1s$
and $2s$ levels a few different transitions such as $1s-2s$,
$1s-3s$, $2s-6s/d$, $2s-8s/d$ and $2s-12d$ transitions
\cite{Beauvoir} were used. The accuracy of the latter transitions
with the exception of the $1s-2s$ transition \cite{Niering} is not
so high as for the theoretical one. So, having more precisely
measured frequencies of the above mentioned transitions one can
obtain more accurate values for the Lamb shift of $1s$ and $2s$
levels in hydrogen atom and, as a consequence, this leads to a
more precise estimation of minimal length.

It was already mentioned that the estimation of the minimal length
can be received if we make use of the assumption:
\begin{equation}
\Delta^{\textrm{ml}}_2\leq
\Delta^{\textrm{exprt}}_2-\Delta^{\textrm{theor}}_2
\end{equation}
where $\Delta^{\textrm{ml}}_2=8\Delta E^{(1)}_{2s}-\Delta
E^{(1)}_{1s}$ is a special difference constructed similarly to
(\ref{difference}) and $\Delta E^{(1)}_{1s}$, $\Delta
E^{(1)}_{2s}$ are the corrections to the $1s$ and $2s$ energy
levels caused by deformation of commutation relations.

The difference $\Delta^{\textrm{ml}}_2$ can be represented in the
following form:
\begin{equation}\label{diff2}
\Delta^{\textrm{ml}}_2=\frac{e^2\hbar^2}{a^3}\left(\frac{1}{2}(\beta+\beta')-(2\beta-\beta')
\left(\frac{3}{2}-\ln(4)\right)\right).
\end{equation}

Similarly to \cite{benczik, mykola} we introduce two dimensionless
parameters $\xi=\frac{\Delta x_{min}}{a}$ and
$\eta=\frac{\beta}{\beta+\beta'}$ instead of $\beta$ and $\beta'$,
where the minimal length $\Delta
x_{min}=\hbar\sqrt{\beta+\beta'}$. As was noted in \cite{mykola}
our calculations take place if $2\beta-\beta'\geq 0$ and $\beta$,
$\beta'$ are nonnegative constants. So we have the constraints on
the domain of variation for the dimensionless parameter $\eta$:
$\frac{1}{3}\leq\eta\leq 1$. We rewrite the right hand side of
expression (\ref{diff2}) using the parameters $\eta$ and $\xi$:
\begin{equation}\label{diff3}
\Delta^{\textrm{ml}}_2=\frac{e^2}{a}\xi^2\left(\frac{1}{2}-(3\eta-1)\left(\frac{3}{2}-\ln(4)\right)\right).
\end{equation}

It is easy to obtain the simple expression for the estimation of
the minimal length using the relation (\ref{diff3}):
\begin{equation}\label{minlength}
\Delta x_{min}=\xi
a=a\sqrt{\frac{a}{e^2}\,\frac{2\Delta^{\textrm{ml}}_2}{1-(3\eta-1)(3-2\ln(4))}}
\end{equation}
So if we suppose that $\Delta^{\textrm{ml}}_2$ is equal to the
difference $\Delta^{\textrm{exprt}}_2-\Delta^{\textrm{theor}}_2$
we can numerically calculate the minimal length for an arbitrary
parameter $\eta$ on its domain of variation.

\begin{figure}[hb!!]
\centerline{\includegraphics[scale=0.9,clip]{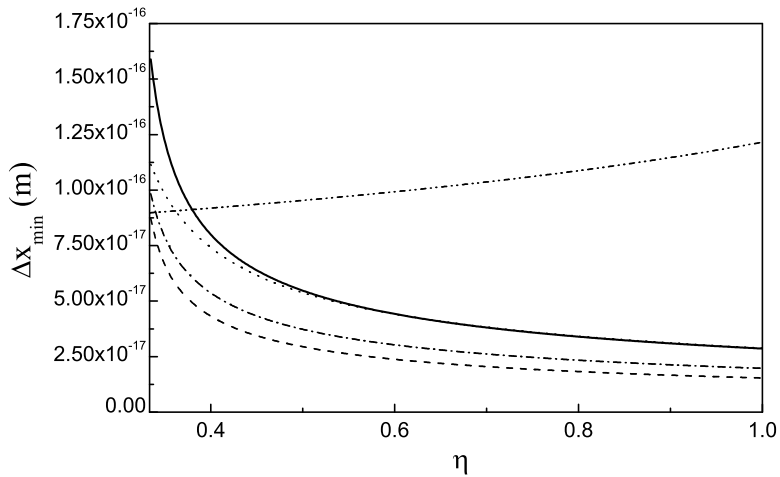}}
\caption{The comparison of the estimations for the minimal length
obtained by using different approaches. The solid and the dotted
lines represent the estimations obtained in the works
\cite{mykola} and \cite{benczik} respectively by using the same
experimental and theoretical data for the $1s$ Lamb shift. The
dashed line shows the constraints on the minimal length obtained
in the work \cite{mykola} by using more recent data
\cite{Beauvoir, pachucki} for the $1s$ Lamb shift. The dash-dotted
line shows the estimation for the minimal length obtained by using
the data \cite{Beauvoir, pachucki} for the $2s$ Lamb shift. At
last the dash-dot-dotted line represents the constraints on the
minimal length that were obtained by using the expression
(\ref{minlength}). }\label{rys2}.
\end{figure}

We also compare the constraints on the minimal length with the
results obtained in \cite{benczik, mykola}. This comparison is
represented in Fig.\ref{rys2}. As it is easy to see the behavior
of the minimal length obtained with using expression
(\ref{minlength}) is qualitatively different from the preceding
estimations. So if we enlarge the parameter $\eta$ the minimal
length increases. In the previous investigations the minimal
length decreased with increasing parameter $\eta$. It is necessary
to say that for the evaluation of minimal length we introduced a
special expression $\Delta_2=8\Delta_L(2s)-\Delta_L(1s)$ in
contrast to the previous works \cite{benczik,mykola} where for the
estimation of minimal length the Lamb shift of $1s$-level was
used. So, it is not strange that dependence of minimal length on
the parameter $\eta$ obtained in this paper is somewhat different
from the behavior of minimal length received in
\cite{benczik,mykola}. But we want to stress that the behavior of
the minimal length as a function of the parameter $\eta$ is not so
important as the order of magnitude. We see that estimations of
the minimal length obtained with using two different approaches
give us the minimal length of the same order as it has to be.
\section{Discussion}

We investigated the hydrogen atom problem with the deformed
Heisenberg algebra leading to the existence of minimal length. In
our previous work \cite{mykola} we put forward an effective
perturbation theory giving a possibility to calculate corrections
for arbitrary levels of hydrogen atom including $s$-levels. We
used this perturbation theory and calculated corrections to
$ns$-level for arbitrary $n$. It is necessary to note that in
\cite{mykola} only the corrections to the $1s$ and $2s$ levels
were calculated.

For the estimation of the minimal length we introduced a special
difference $8\Delta_L(2s)-\Delta_L(1s)$. Such a difference gave us
the possibility to obtain a simple relation for the estimation of
minimal length. We stress that our evaluation of the minimal
length does not give stringent result but rather the upper bound
for the minimal length. Comparison of the results obtained here
with experimental data from the precision hydrogen spectroscopy
shows that the upper bound for the minimal length is of the order
$10^{-16}$ m. The behavior of our estimation of minimal length is
somewhat different from the previous one \cite{benczik, mykola}.
As it was noted above the behavior of minimal length is not
important for our estimation. The results obtained here show that
the minimal length has the same order as in the previous works
\cite{benczik,mykola} and this is most important.

\section{Acknowledgments}
The author is grateful to Prof. V. M. Tkachuk for many useful
discussions and comments. I would also like to thank Dr. A. A.
Rovenchak  for a careful reading of the manuscript.

\end{document}